\documentclass[pre,twocolumn,showpacs]{revtex4-1}

\newcommand{\Eq}[1]{Eq.~(\ref{eq:#1})}
\newcommand{\Eqs}[1]{Eqs.~(\ref{eq:#1})}
\newcommand{\Fig}[1]{Fig.~\ref{fig:#1}}
\newcommand{\Figure}[1]{Figure~\ref{fig:#1}}

\newcommand{\Ref}[1]{Ref.~\cite{#1}}

\newcommand{\rr}{\mathbf{r}}
\renewcommand{\v}{\mathbf{v}}

\newcommand{\gdot}{\dot{\gamma}}

\newcommand{\taurelax}{\tau_\mathrm{relax}}
\newcommand{\taudiss}{\tau_\mathrm{diss}}
\usepackage{graphicx}

\begin{document}
\title{Relaxation and Rheology in Dense Athermal Suspensions}

\author{Peter Olsson}

\affiliation{Department of Physics, Ume\aa\ University, 
  901 87 Ume\aa, Sweden}



\date{\today}   

\begin{abstract}
  We study relaxation and rheology of dense athermal suspensions of frictionless
  particles close below the jamming density. Our key quantity, the relaxation
  time---determined from the exponential decay of the energy after the shearing
  has suddenly been switched off---is argued to be a determining factor behind
  the algebraic divergence of various quantities as the jamming density is
  approached from below. We also define and measure the ``dissipation time'',
  which is obtained directly in shearing simulations and find that it behaves
  similarly to the relaxation time. Comparing shear viscosity with the
  expression for the dissipation time we identify a non-divergent factor that
  explains the need for correction terms in the scaling analyses of the shear
  viscosity.
\end{abstract}

\maketitle

As the volume fraction increases in zero-temperature collections of spherical
particles with repulsive contact interaction, there is a transition from a
liquid to an amorphous solid state---the jamming transition. This transition has
for quite some time been studied through simulations in two different ways: by
examining static packings generated by compressing and relaxing random packings,
and by driving the system with a shear deformation. Whereas it was first
commonly expected that these two approaches would show the same behavior, the
evidence now suggest that they are clearly different. One example is the
difference in the behavior of the pressure above the jamming density, $\phi_J$,
$p(\phi)\sim (\phi-\phi_J)^y$, which is linear, $y=1$ for static
packings\cite{OHern_Langer_Liu_Nagel:2002} but appears to be $y\approx1.1$ for
the shear-driven case\cite{Olsson_Teitel:gdot-scale}. Another example is the
isolated mode in the spectrum that dominates the behavior of the shear-driven
system close to the transition\cite{Lerner-PNAS:2012} but which is not present
in static packings.

One way to study the shear-driven transition is to try and eliminate the
complications related to the softness of the particles and instead try and
determine the behavior of hard particles. This is usually done by driving with
sufficiently low shear rates, $\gdot$, such that the particle overlaps become
negligable---this is the linear region where many quantities are linear in
$\gdot$ (see e.g.\ Fig.~1 in \Ref{Olsson_Teitel:jam-HB}). This is so since in
the strict hard core limit one expects particles driven with different $\gdot$,
to follow the same path through phase space, only with different velocities
$\v_i\propto\gdot$, and it then follows that many quantities (e.g.\ the forces)
are just proportional to $\gdot$\cite{Olsson_Teitel:jam-HB, Andreotti:2012}. The
alternative is a recently deviced method to perform shearing simulations with
hard particles\cite{Lerner-PNAS:2012, Lerner-Comp:2013}.

The transition in shear-driven systems still appears to be rather poorly
understood. There is e.g.\ no accepted value for the exponent for the divergence
of the viscosity; determined values range between 2.0 and
2.8\cite{Olsson_Teitel:jamming, Hatano:2008, Otsuki_Hayakawa:2009b,
  Bonnoit:2010, Olsson_Teitel:gdot-scale, Boyer:2011, Andreotti:2012}, and this
appears to, to at least some extent, be because of the lack of understanding of
the mechanism behind this divergence. To illustrate the complications we point
out that one typically expects both shear viscosity $\eta=\sigma/\gdot$ and the
pressure-equivalent quantity, $\eta_p=p/\gdot$, to diverge in the same way, but
since $\mu=\sigma/p=\eta/\eta_p$ has a pronounced density
dependence\cite{Boyer:2011,Lerner-PNAS:2012} in the relevant density interval,
naive fits of $\eta(\phi)$ and $\eta_p(\phi)$ to algebraic divergences,
$(\phi_J-\phi)^{-\beta}$, give differing values for the critical parameters. One
way to resolve this issue is to include corrections to scaling in the analyses,
but even though such a program has been successfully
accomplished\cite{Olsson_Teitel:gdot-scale}, this requires very high precision
data very close to the transition and the scaling analysis becomes both
difficult and opaque.

In this Letter we present results from relaxation simulations which are done by
first driving at a certain shear rate and then stopping the shearing and letting
the system relax according to its dynamics. The relaxation time $\taurelax$ is
then determined from the decay of the energy. We believe that this relaxation
time is a fundamental quantity which is at the root of the divergence of
pressure and shear viscosity. We also consider another time, $\taudiss$, which
is related to the rate at which energy is dissipated in steady shearing and show
that this quantity behaves similarly to $\taurelax$. We further show that $\eta$
and $\eta_p$ may be written as products of $\taudiss$ and some $\phi$-dependent
correction factors, and we show that this picture nicely explains the need for
corrections to scaling in the scaling analysis of
\Ref{Olsson_Teitel:gdot-scale}. The methods suggested here should be useful for
studies of the jamming transition through both simulations and experiments.

We simulate frictionless soft disks in two dimensions using a bi-dispersive
mixture with equal numbers of disks with two different radii of ratio
1.4. Length is measured in units of the diameter of the small particles
($d_s=1$). With $r_{ij}$ the distance between the centers of two particles,
$d_{ij}$ the sum of their radii, and the relative overlap $\delta_{ij}=1 -
r_{ij}/d_{ij}$ for $r_{ij}<d_{ij}$ and $\delta_{ij}=0$ otherwise, the
interaction between the particles is
\begin{displaymath}
  V(r_{ij}) = \epsilon \delta_{ij}^2/2,
\end{displaymath}
with $\epsilon=1$.  We use Lees-Edwards boundary conditions \cite{Evans_Morriss}
to introduce a time-dependent shear strain $\gamma(t) = t\gdot$. With periodic
boundary conditions on the coordinates $x_i$ and $y_i$ in an $L\times L$ system,
the position of particle $i$ in a box with strain $\gamma$ is defined as $\rr_i
= (x_i+\gamma y_i, y_i)$.  We simulate overdamped dynamics at zero temperature
with the equation of motion \cite{Durian:1995},
\begin{displaymath}
  \frac{d\rr_i}{dt} = -\frac{1}{k_d}\sum_j\frac{dV(\rr_{ij})}{d\rr_i} + y_i \gdot\; \hat{x},
\end{displaymath}
with $k_d=1$. In this model dissipation occurs when the particles move relative
to the steady shearing velocity $y_i \gdot \hat{x}$. The effects of instead
letting the dissipation be given by the relative velocity of particles in
contact will be discussed elsewhere\cite{Vagberg_Olsson_Teitel:NMF}.

\begin{figure}
  \includegraphics[width=6cm]{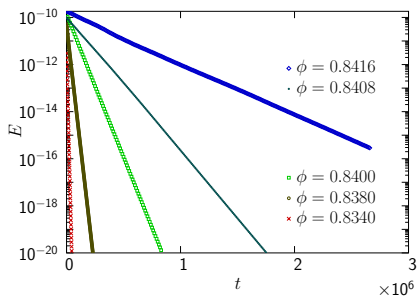}
  \caption{Relaxation of the energy at different $\phi$. The figure shows the
    relaxation of energy after the shearing has been switched off.  The
    preceeding shearing simulations were performed at very low shear rates in
    order to stay close to the linear region; the densities and the initial
    shear rates were $(\phi,\gdot)=(0.834, 10^{-8})$, $(0.838, 10^{-8})$,
    $(0.840, 5\times10^{-9})$, $(0.8408, 2\times10^{-9})$, $(0.8416, 10^{-9})$.
    To determine the relaxation times, $\taurelax$, we fit the energy to an
    exponential decay, only using data with $E<10^{-12}$.}
  \label{fig:E-time}
\end{figure}

The key quantity in this Letter, the relaxation time, is determined through a
two-step process: first the system is driven in steady shear at a constant shear
rate $\gdot$; then the shearing is stopped and the system is allowed to relax
down to a minimum energy. As the simulations discussed here are at densities
somewhat below $\phi_J$, the final state is always a state of zero energy, and
after a short transient time, the decay is exponential,
\begin{displaymath}
  E(t) \sim \exp(-t/\taurelax).
\end{displaymath}
A few realisations of such relaxations are shown in \Fig{E-time}. For each
realisation the relaxation time is determined from the data with $E(t)<10^{-12}$,
where the decay is exponential to an excellent approximation. We determine
$\taurelax(\phi,\gdot)$ as the average relaxation time from about 10--100 such
relaxations.

\begin{figure}
  \includegraphics[bb=51 324 334 654, width=4cm]{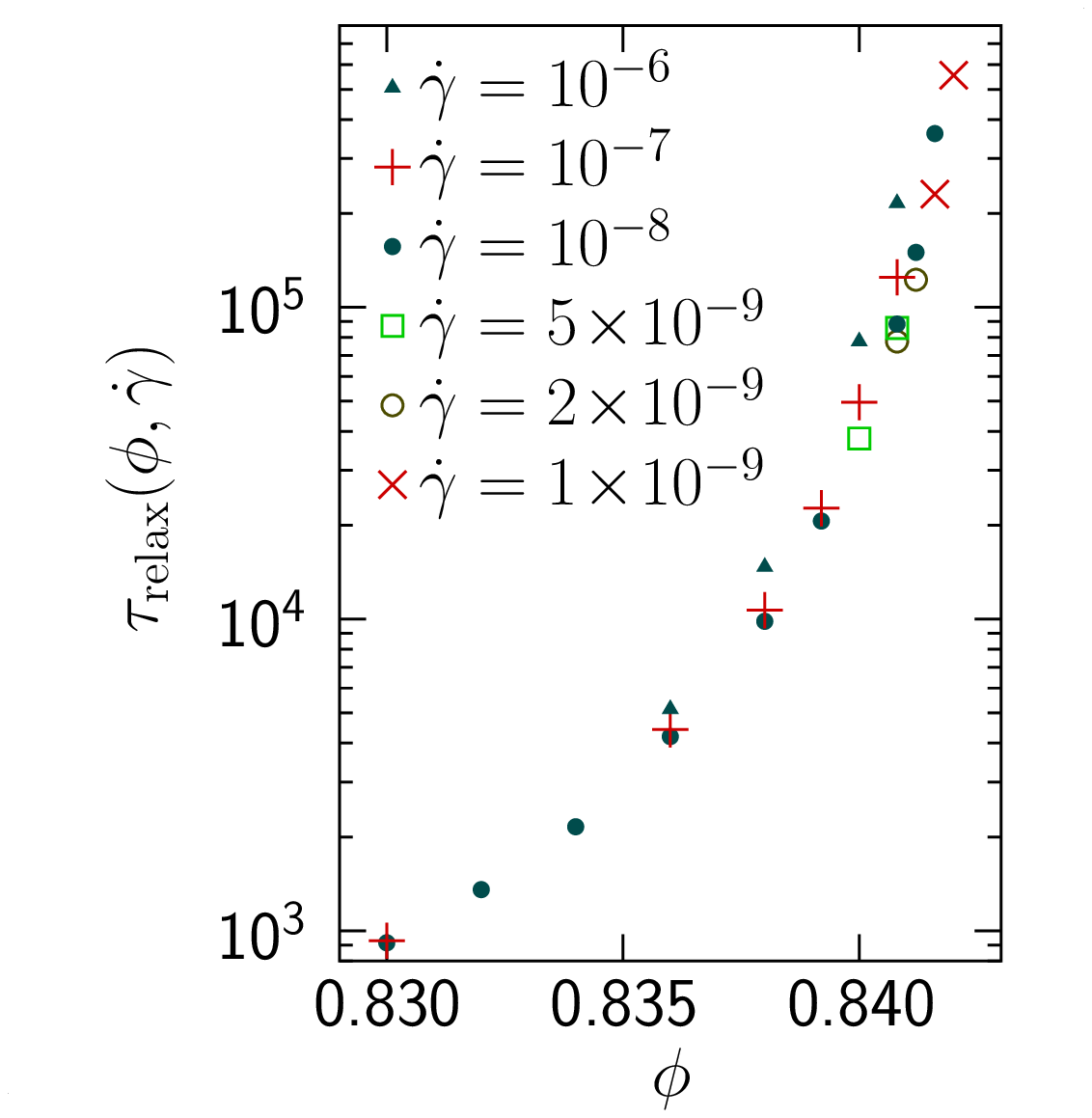}
  \includegraphics[bb=51 324 334 654, width=4cm]{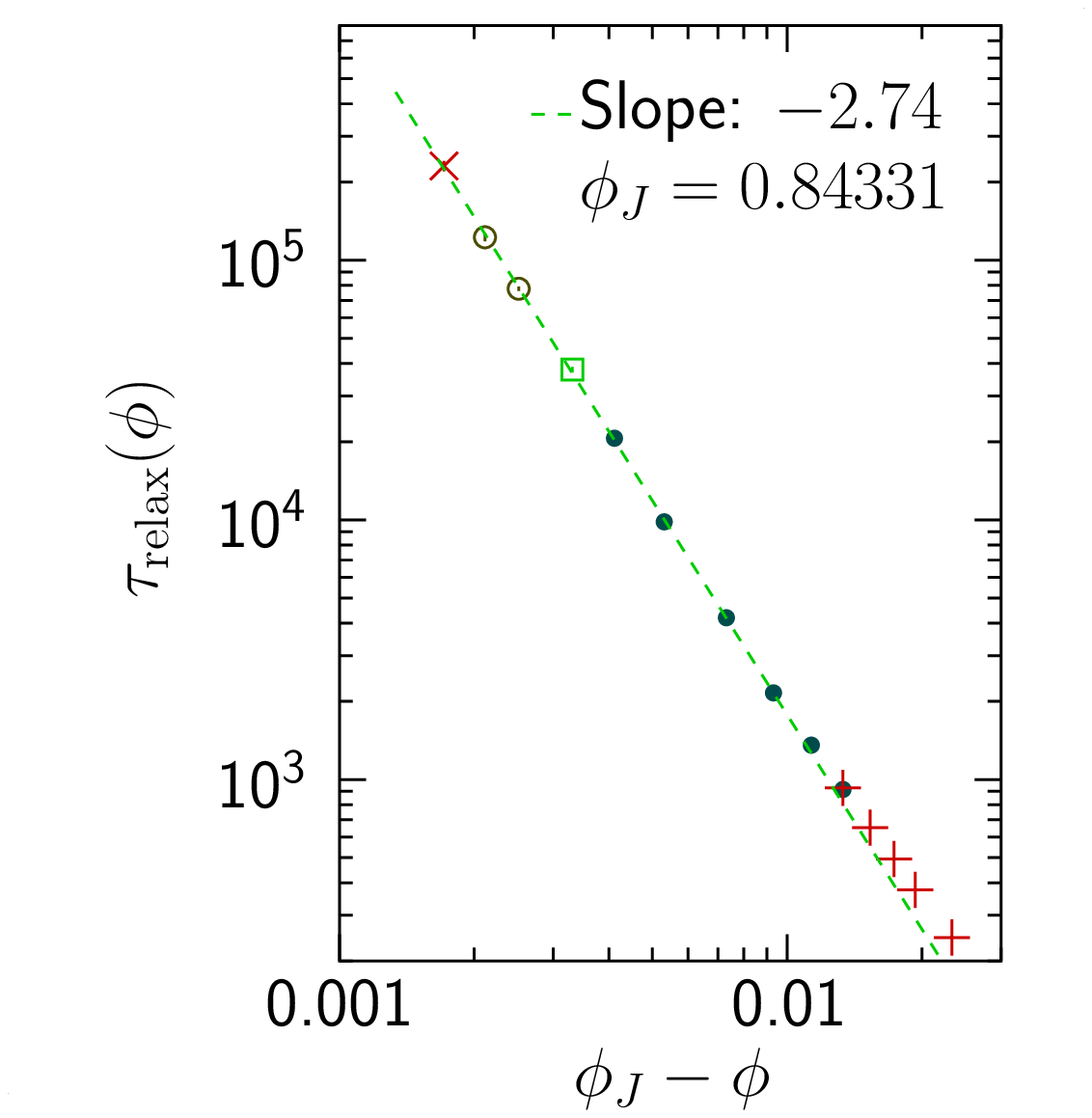}
  \caption{Behavior of the relaxation time. Panel (a) shows how $\taurelax$
    depends on both $\phi$ and $\gdot$, which is here the shear rate of the
    preparatory run (the relaxations are performed with $\gdot=0$). Each value
    here is the average of relaxation times determined from of a large number of
    different relaxations. At sufficiently low $\gdot$, $\taurelax$ approaches
    well-defined values that only depend on $\phi$. Panel (b) is a double-log
    plot, only including the points with small enough $\gdot$ to be in the linear
    region. The figure shows a fit to the eight points with $\phi\geq 0.834$;
    the points with $\phi_J-\phi>0.01$ are not included in the fit.}
  \label{fig:taurelax}
\end{figure}

\Figure{taurelax}(a), which is $\taurelax$ versus $\phi$ in a narrow density
interval just below $\phi_J$, shows that $\taurelax$ increases very rapidly with
$\phi$. The data are shown for a few different $\gdot$ and we conclude that
$\taurelax$ at a given $\phi$ approaches a well-defined limiting value as
$\gdot$ decreases and the linear region is approached. In this Letter we analyze
the data within (or close to) this linear region only; the behavior at larger
$\gdot$ will be examined elsewhere.  We first determine the critical behavior
from the eight points in \Fig{taurelax}(a) which are in the linear region and
close below jamming, i.e.\ the points with the lowest shear rate for each
density in the range $0.834\leq\phi\leq0.8416$. Fitting these points to an
algebraic divergence, $\taurelax(\phi) = A|\delta\phi|^{-\beta}$, (where
$\delta\phi=\phi_J-\phi$) gives \Fig{taurelax}(b) and the critical parameters
$\phi_J=0.8433\pm0.0002$ and $\beta=2.73\pm0.15$, which are in good agreement
with Refs.~\cite{Heussinger_Barrat:2009,Olsson_Teitel:gdot-scale}. The quoted
errors represent max/min-values, corresponding to three standard deviations in
the estimated quantities.

Our assumption is that this increase of the relaxation time as jamming is
approached is the fundamental phenomenon which is at the root of the divergence
of other quantities as e.g.\ the shear viscosity, $\eta$. The relaxation mode
should be related to the isolated mode with frequency $\omega_\mathrm{min}$ in
\Ref{Lerner-PNAS:2012}, and we expect
$\taurelax\sim\omega_\mathrm{min}^{-2}$. (The different powers of time here
reflect the differences in dynamics. In overdamped dynamics there is a velocity
that is proporional to a force, whereas one in vibrational analyses assumes
Newtonian dynamics with massive particles where the acceleration is proportional
to the force.)  Note also that the lowest mode being \emph{isolated} explains
why the relaxation is almost perfectly exponential after the initial decay. One
would otherwise typically expect $E(t)$ to be given by a sum of several modes
with close but different time constants. A further result from
\Ref{Lerner-PNAS:2012, Lerner-epl:2012} is that $\omega_\mathrm{min}^{-2}$ (and
thereby $\taurelax$) diverges with the same exponent as the shear
viscosity. This conclusion will also be reached below in a different way.

We will now try and establish a link between the shear viscosity, $\eta$, which
is typically measured in both simulations and experiments, and the above
obtained $\taurelax$. This will be done in two steps: we first derive an
expression for a similar time, $\taudiss$, which is obtained directly in the
shear driven simulations; we then express $\eta$ in terms of $\taudiss$.

The expression for the dissipation time, $\taudiss$, is obtained from a power
balance. The idea is that the supplied power, which is $\sigma\gdot$ per unit
area, on average should be balanced by the dissipated power. Defining $\taudiss$
such that $E/\taudiss$ is the rate at which the energy is dissipated defines
\begin{equation}
  \taudiss = \frac{E}{\sigma\gdot}.
  \label{eq:taudiss}
\end{equation}
Note that it follows directly that $\taudiss$ should diverge with the exponent
$\beta$ since $\sigma/\gdot \sim |\delta\phi|^{-\beta}$ and $E/\gdot^2 \sim
|\delta\phi|^{-2\beta}$\cite{Olsson_Teitel:jam-HB}.

\begin{figure}
  \includegraphics[bb=51 324 334 654, width=4cm]{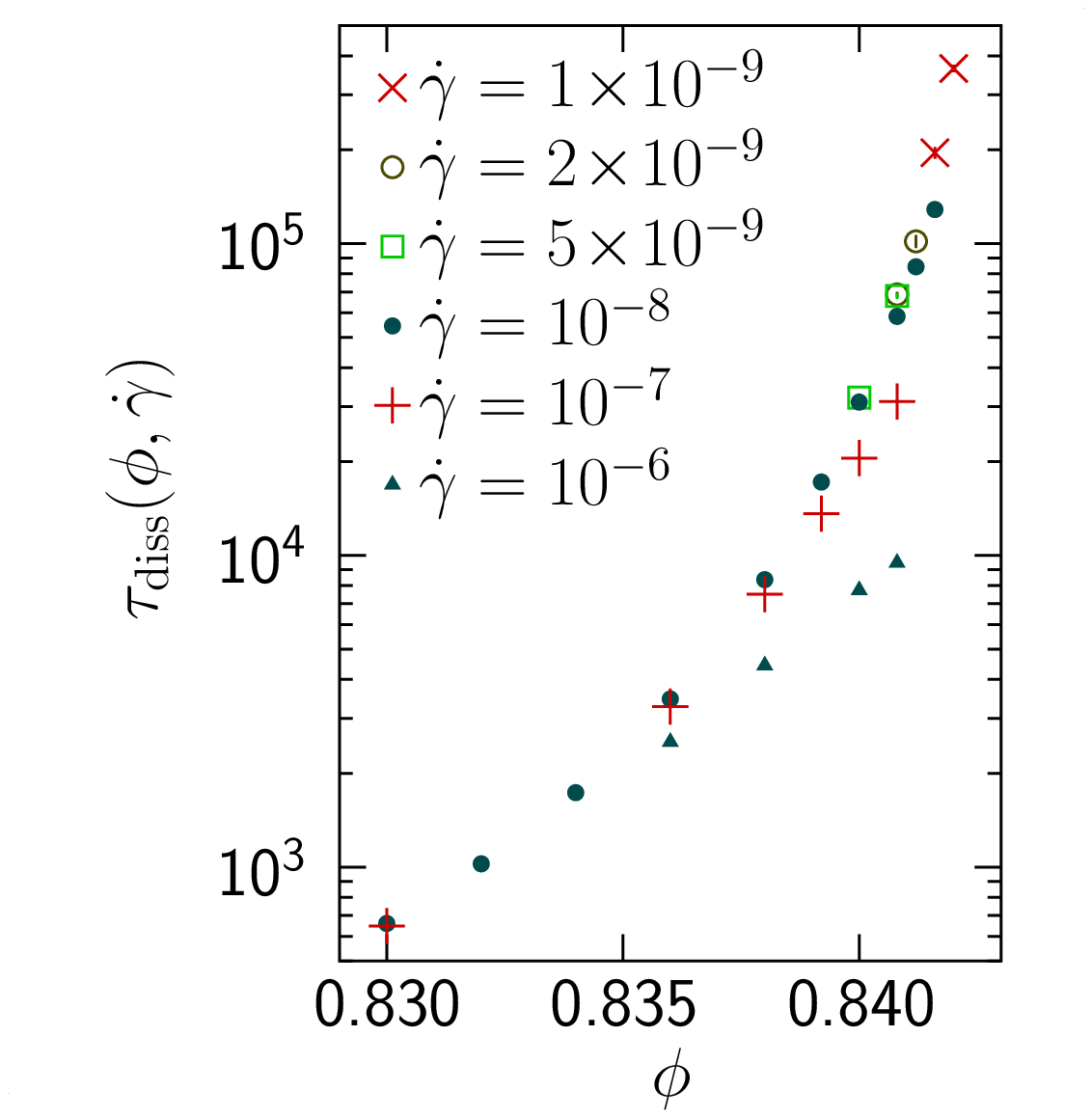}
  \includegraphics[bb=51 324 334 654, width=4cm]{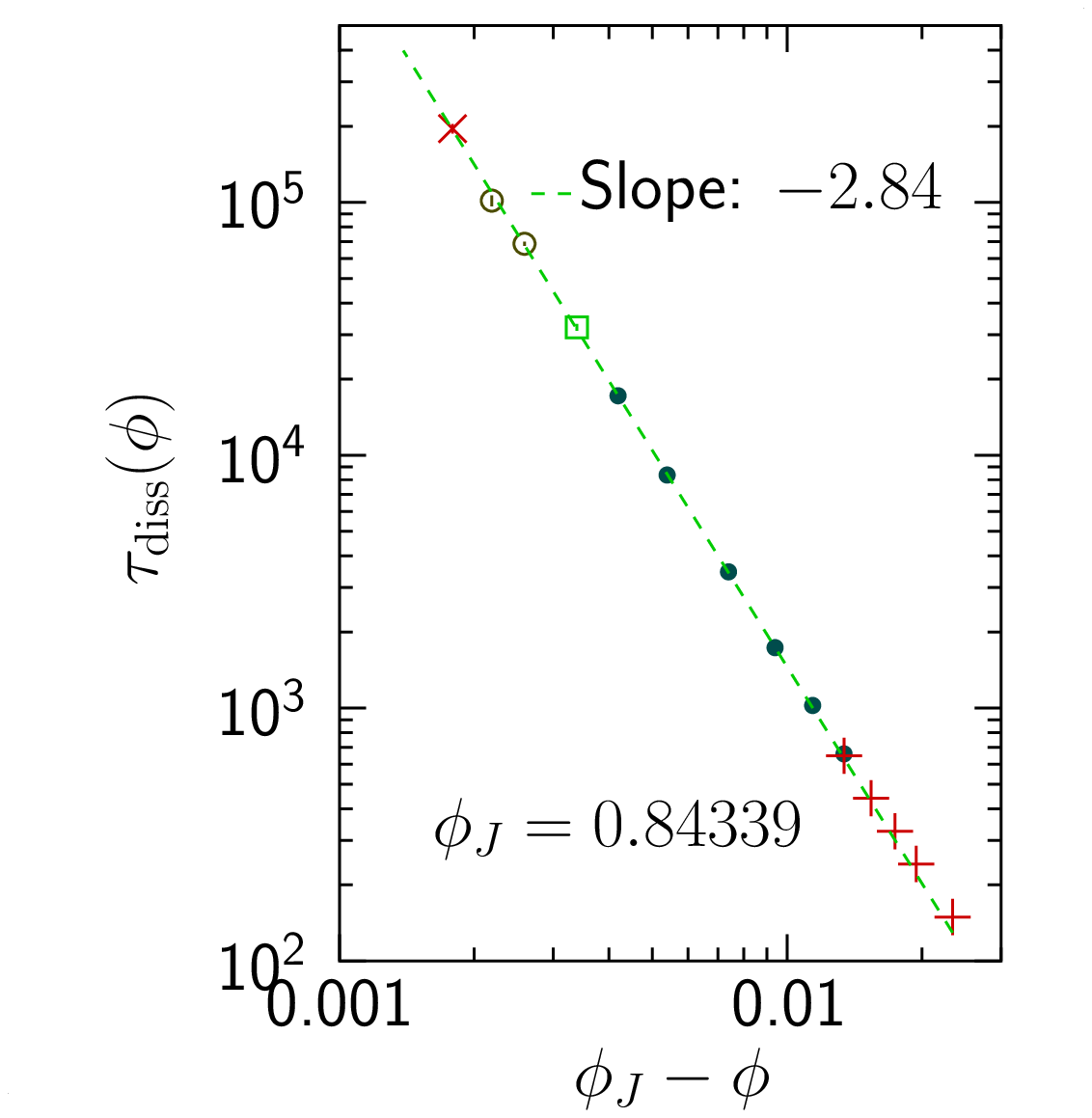}
  \caption{Dissipation time $\taudiss$ obtained in shear driven
    simulations. Panel (a) is $\taudiss$ against $\phi$ at different
    $\gdot$. Panel (b) shows the data considered to be in the linear region
    against $\phi_J-\phi$. The dashed line shows the result of a fit to the
    eight points with $\phi_J-\phi<0.01$, giving $\phi_J=0.8434$ and
    $\beta=2.84$.}
  \label{fig:taudiss}
\end{figure}

The dissipation time $\taudiss$ versus $\phi$ for a few different shear rates is
shown in \Fig{taudiss}(a). Just as for $\taurelax$ we find that $\taudiss$
increases rapidly when $\phi$ increases towards $\phi_J$, and we also find
well-defined low-$\gdot$ limits, with deviations for larger $\gdot$. Here
$\taudiss$ becomes \emph{smaller} for larger $\gdot$, which is the same behavior
as in the shear viscosity, but opposite to the behavior of $\taurelax$,
discussed above.

\begin{figure}
  \includegraphics[width=7cm]{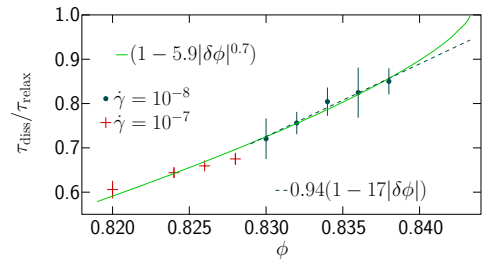}
  \caption{Relation between dissipation time and relaxation time. The figure
    shows that $\taudiss/\taurelax$ increases slowly as $\phi_J$ is approached
    from below.  The solid line shows a fit to an algebraic function which
    approaches unity at $\phi_J$. The dashed line is a parametrization to help
    compare the size of this correction with the other corrections to scaling in
    \Fig{eta-etae-dphi}.}
  \label{fig:taudiss-taurelax}
\end{figure}

The rational to introduce $\taudiss$ was to find a quantity in the shearing
simulations that behaves similarly to $\taurelax$, and thus establish a link
between the relaxation dynamics and the shearing simulations. It is however
clear that these two quantities cannot be identical. Since the initial
dissipation in a relaxation simulation has to be the same as the dissipation
under steady shear, $\taudiss$ is equal to the \emph{initial} decay rate in a
relaxation simulation. $\taurelax$ on the other hand is the decay rate at long
times. This means that $\taudiss$ should get contributions from all decay modes
that are present in the system. $\taurelax$, on the other hand, is determined by
the slowest mode only, since that is the only mode that persists after
sufficiently long times.  Since $\taudiss$ gets contributions from modes with
smaller time constants it follows that $\taudiss<\taurelax$. This is confirmed
by \Fig{taudiss-taurelax} which, furthermore, shows that $\taudiss/\taurelax$
increases with increasing $\phi$ and appears to approch unity as
$\phi\to\phi_J$. We relate this to the observation in \Ref{Lerner-PNAS:2012}
that the relative contribution to the shear viscosity of the isolated mode (in
their notation, $\sigma_0/\sigma$) approaches unity as jamming is approached,
which means that the weight of the other modes decreases. We likewise expect
the contributions from the faster modes to $\taudiss$ to become less important
as $\phi_J$ is approached, which implies $\taudiss/\taurelax\to1$.  We summarize
the above in terms of two conclusions of importance for the present work: (i)
Properties determined in steady shear will necessarily be different from the
properties determined from the long-time behavior of the relaxation
simulations. (ii) This difference is however rather small and one should
therefore expect results based on $\taurelax$ and $\taudiss$, respectively, to
be very similar.

A fit of $\taudiss$ to the algebraic divergence (where we again use only the
eight points in the linear region and close to $\phi_J$) is shown in
\Fig{taudiss}(b) and gives $\phi_J=0.8434\pm0.0003$ and
$\beta=2.84\pm0.20$. Both values are close to (just slightly higher than) the
corresponding values from the analysis of $\taurelax$, and this again suggests
that $\taudiss$ is a good approximation of $\taurelax$.

\begin{figure}
  \includegraphics[bb=51 324 334 654, width=4cm]{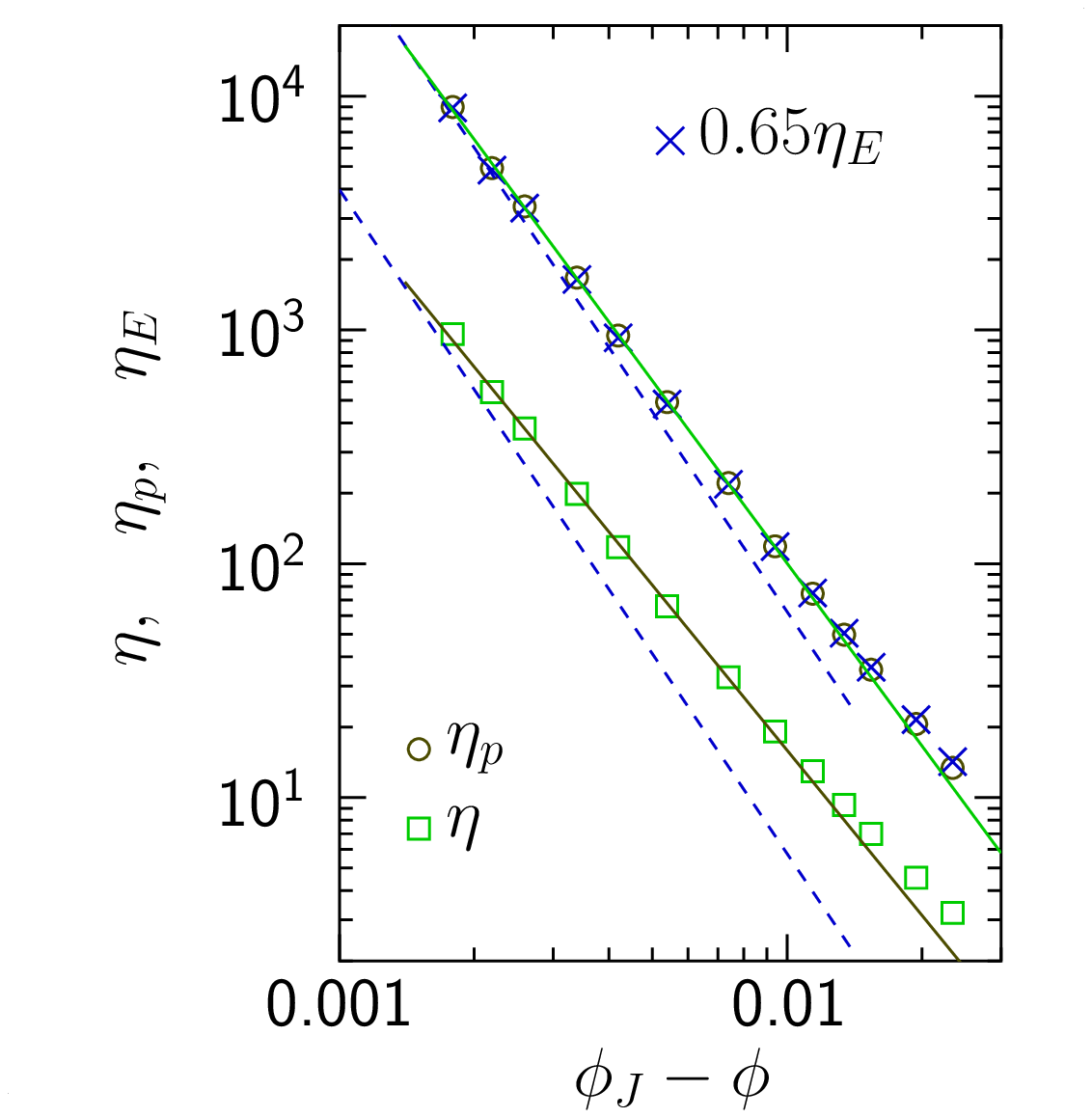}
  \includegraphics[bb=51 324 334 654, width=4cm]{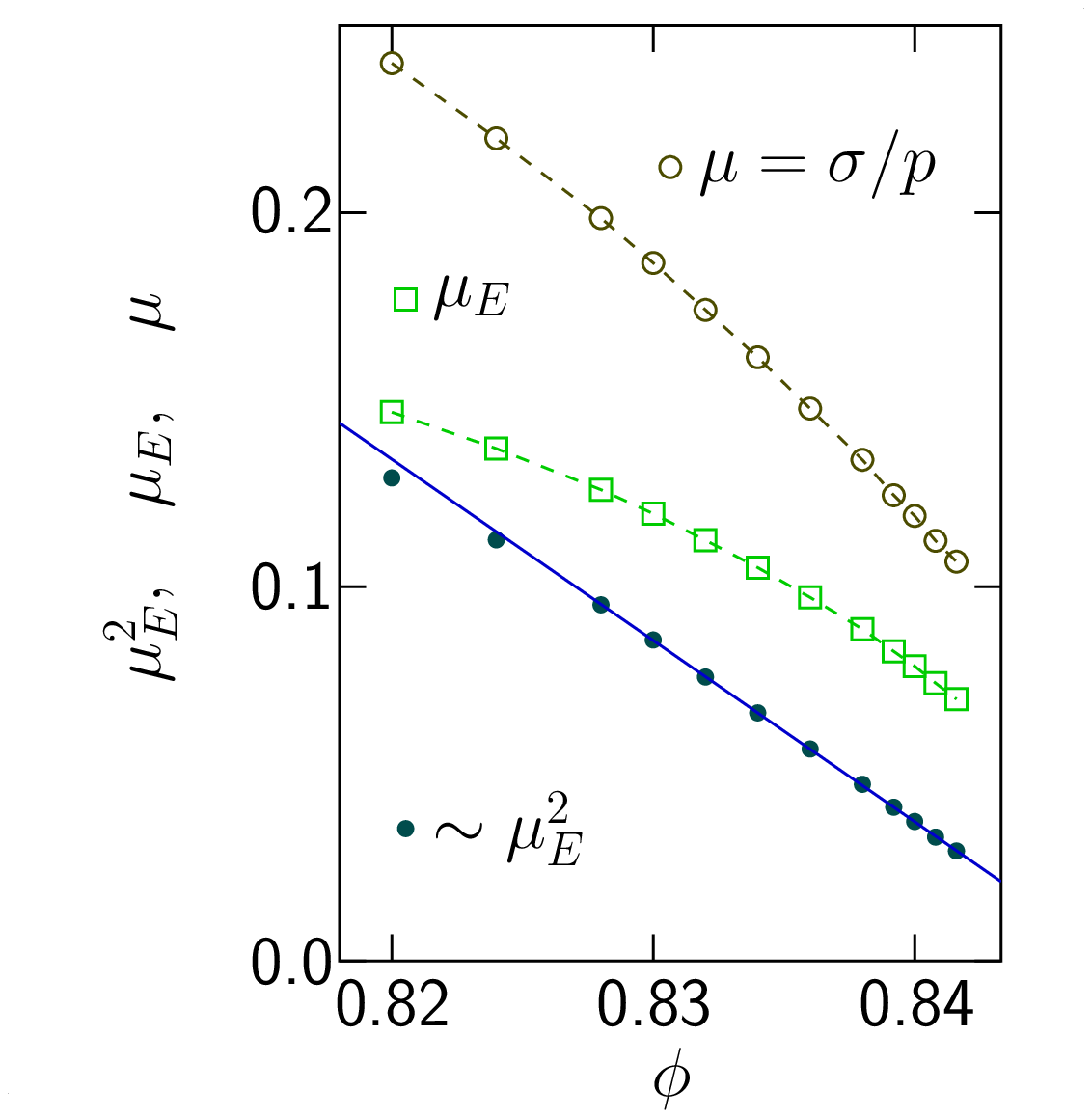}
  \caption{Viscosities $\eta$, $\eta_p$, and $\eta_E$ in the light of \Eqs{eta}
    and (\ref{eq:etae}). The dashed lines in panel (a) show the divergence of
    $\taudiss$ with $\beta=2.84$.  The symbols show how $\eta$ and $\eta_p$
    approach the presumed asymptotic scaling behavior, and it is clear that the
    corrections to scaling are much larger for $\eta$ than for $\eta_p$. The
    same is seen by naively fitting $\eta$ and $\eta_p$ to algebraic divergences
    (shown by solid lines) which give $\beta=2.35$ and $2.59$, respectively,
    where we note that the exponent obtained for $\eta$ is further off the
    asymptotic value $\beta=2.84$.  Panel (b) shows both
    $\mu_E=\sigma/\sqrt{E}$, the similar dimensionless friction $\mu=\sigma/p$,
    and $\mu^2_E$ (the correction in \Eq{eta}). This last quantity appears to be
    linear in $\phi$, which is consistent with $\omega\nu\approx1$ as found in
    \Ref{Olsson_Teitel:gdot-scale}.}
  \label{fig:eta-etae-dphi}
\end{figure}

The finding that $\taudiss$ to a good approximation diverges algebraically,
gives a ground for understanding the need for corrections to scaling in the
analyses of $\eta$ and $\eta_p$ in \Ref{Olsson_Teitel:gdot-scale}.  For the
$\gdot\to 0$ limit at densites below $\phi_J$, corrections to scaling means that
the divergence cannot be well approximated by the algebraic
$A|\delta\phi|^{-\beta}$ alone, but that one instead has to use
\begin{equation}
  A|\delta\phi|^{-\beta} \left(1 + a |\delta\phi|^{\omega\nu}\right),
  \label{eq:corr}
\end{equation}
which follows from using $b=|\delta\phi|^{-\nu}$ in the unnumbered equation
before Eq.~(3) in \Ref{Olsson_Teitel:gdot-scale}. Here the correction to scaling
exponent $\omega$ appears together with the correlation length exponent $\nu$.
This behavior is illustrated in \Fig{eta-etae-dphi}(a) which shows both $\eta$
and $\eta_p$ versus $\phi_J-\phi$. Also shown is $\eta_E=\sqrt{E}/\gdot$ which
behaves the same as $\eta_p$, to an excellent approximation. (This is so since
$p\sim\sum_{ij} \delta_{ij}$ whereas $E\sim
\sum_{ij}\delta^2_{ij}$\cite{Hatano:2009}.)
As is clear from the figure, $\eta$ and $\eta_p$ behave differently, and
attempts to determine $\beta$ from algebraic fits without corrections, give
$\beta=2.35$ and $2.59$, respectively, as shown by the solid lines. Since one
expects the asymptotic behavior of $\eta$ and $\eta_p$ to be the same, this
discrepancy calls for including corrections to scaling (as in \Eq{corr}), which
was also done successfully in \Ref{Olsson_Teitel:gdot-scale}.

We will now show that $\eta$ and $\eta_E$ may be written as products of
$\taudiss$ and some correction factors. After introducing $\mu_E =
\sigma/\sqrt{E}$ in analogy with the dimensionless friction $\mu=\sigma/p$ (see
\Fig{eta-etae-dphi}(b)) we find using \Eq{taudiss} that
\begin{eqnarray}
  \label{eq:eta}
  \eta & = & \sigma/\gdot = 
  \mu_E^2\; \taudiss, \\
  \label{eq:etae}
  \eta_E & = & \sqrt{E}/\gdot = 
  \mu_E\; \taudiss.
\end{eqnarray}
We note two things: (i) That the corrections of $\eta$ and $\eta_E$ are
$\mu_E^2$ and $\mu_E$, respectively gives a very direct explanation to why the
correction to scaling in $\eta_p$ (which behaves essentially the same as
$\eta_E$) is so much smaller than in $\eta$\cite{Olsson_Teitel:gdot-scale}. See
also below for a direct comparison of these correction terms. (ii) As shown in
\Fig{eta-etae-dphi}(b) $\mu^2_E$ (the correction factor that goes together with
$\eta$) is linear in $\phi$ to an excellent approximation and the same holds for
$\mu_E$, though in a more narrow range below $\phi_J$. Together with \Eq{corr}
we therefore conclude that $\omega\nu\approx1$, again in agreement with
\Ref{Olsson_Teitel:gdot-scale}.

As discussed above our starting assumption is that $\taurelax$ diverges
algebraically and it then follows from \Fig{taudiss-taurelax} that $\taudiss$ is
given by this algebraic divergence times a correction factor. To argue that the
critical behavior should be determined from $\taudiss$ rather than $\eta$ or
$\eta_p$, we now want to show that this correction in $\taudiss$ that one cannot
eliminate (if one only has access to data from steady shearing) is considerably
smaller than the correction factors in $\eta$ and $\eta_E$.  To do that we write
each correction on the form $(1+a|\delta\phi|)$ and compare the magnitude of
``$a$'' for the different cases.  We then find $\mu_E^2 \approx
0.0035(1+228|\delta\phi|)$ and (close to $\phi_J$) $\mu_E\approx
0.061(1+88|\delta\phi|)$ and note that both these correction terms are clearly
bigger than the correction in $\taudiss/\taurelax \sim (1-17|\delta\phi|)$ from
\Fig{taudiss-taurelax}\cite{not-linear}.  This strengthens our confidence in the
use of $\taudiss$ for determining the critical behavior, though it is of course
$\taurelax$ that is the ideal quantity for such analyses.

The results above should also be useful for analyzing experiments, but instead
of using $\taudiss = E/\sigma\gdot$ one could then make use of $\taudiss\approx
p^2/\sigma\gdot$ which is an expression in terms of pressure instead of the
elastic energy. This could be advantageous since pressure should be more readily
available in experiments than energy.

The relaxation dynamics around the jamming transition has been studied before,
but then with a rather different preparation of the starting configurations
\cite{Hatano:2009}. In that study configurations were first generated randomly,
then relaxed to a zero-energy state with the conjugate gradient method, and
after that perturbed by a pure affine shear deformation. The relaxation time was
then determined from the relaxation of such initial states by fitting the shear
stress to $\sigma(\phi,t) \sim t^{-\alpha} e^{-t/\tau}$ with $\alpha=0.55(5)$,
and was found to diverge as $\tau\sim(\phi_J-\phi)^{-\zeta}$ with
$\zeta=3.3(1)$. This exponent is clearly bigger than our $\beta=2.73\pm0.15$. One
possible explanation for this difference is that in the present study we have
been very careful to apply a slow shear driving in the preparation step, whereas
they in their work apply the pure shear deformation suddenly, which should be
more like a rapid shearing. Indeed, as shown in \Fig{taurelax}(b) any given
fixed shear rate would give too large values for $\taurelax$ as one gets close
to $\phi_J$, and from analyses of such data one would expect to get a too high
value of the exponent for the divergence.

To conclude, we have determined $\taurelax$ from relaxational simulations and
suggest that the slowing down of the relaxation as $\phi_J$ is approached is the
fundamental reason for the divergence of $\eta$ and other similar
quantities. Strong support for this idea is obtained from the finding by others
that there is an isolated mode that dominates the behaviour close to
$\phi_J$\cite{Lerner-PNAS:2012}.  We have further introduced $\taudiss$ which is
determined directly in shear driven simulations and have shown that these two
quantities, in the linear region and close to $\phi_J$, are very similar.
From the connection between $\taudiss$ and $\eta$ we further argue that the need
for corrections to scaling in analyses of $\eta$ and related quanties is largely
due to the $\phi$-dependence of $\mu_E=\sigma/\sqrt{E}$. Our results should also
be helpful for getting more accurate determinations of the critical behaviour
from experimental data.

I thank S. Teitel for helpful discussions and critical reading of the
manuscript. This work was supported by the Swedish Research Council grant
2010-3725. Simulations were performed on resources provided by the Swedish
National Infrastructure for Computing (SNIC) at PDC and HPC2N.


\begin{thebibliography}{10}
\bibitem{not-linear}%
  \BibitemOpen
  \bibinfo {note} {The linear relation here is an approximate parametrization
  to help compare the size of the different correction terms. As shown in
  \Fig{taudiss-taurelax} we expect the true behaviour to be given by a
  different exponent.}%
  \bibAnnoteFile{Stop}{not-linear}%
\end{thebibliography}
\end{document}